\documentclass[
superscriptaddress,
twocolumn,
showpacs,preprintnumbers,
amsmath,amssymb,
aps,
prb,
]{revtex4-2}
\usepackage{graphicx}
\usepackage{multirow}
\usepackage{float}
\usepackage[colorlinks,
linkcolor=blue, urlcolor=blue, anchorcolor=blue, citecolor=blue]{hyperref}
\begin{document}

\renewcommand{\tablename}{Table}
\renewcommand{\figurename}{Fig.}
\title{Ferromagnetic half-metallicity in YBaCo$_2$O$_6$ and spin-states driven \\metal-insulator transition}

\author{Wentao Hu}
\affiliation{Laboratory for Computational Physical Sciences (MOE),
 State Key Laboratory of Surface Physics, and Department of Physics,
 Fudan University, Shanghai 200433, China}
\author{Ke Yang}
\affiliation{College of Science,
 University of Shanghai for Science and Technology,
 Shanghai 200093, China}
\affiliation{Laboratory for Computational Physical Sciences (MOE),
	State Key Laboratory of Surface Physics, and Department of Physics,
	Fudan University, Shanghai 200433, China}
\author{Xuan Wen}
\affiliation{Laboratory for Computational Physical Sciences (MOE),
	State Key Laboratory of Surface Physics, and Department of Physics,
	Fudan University, Shanghai 200433, China}
	
\author{Hua Wu}
\email{Corresponding author. wuh@fudan.edu.cn}
\affiliation{Laboratory for Computational Physical Sciences (MOE),
 State Key Laboratory of Surface Physics, and Department of Physics,
 Fudan University, Shanghai 200433, China}
\affiliation{Collaborative Innovation Center of Advanced Microstructures,
 Nanjing 210093, China}

 \date{\today}

\begin{abstract}
Cobaltates have rich spin-states and diverse properties. Using spin-state pictures and first-principles calculations, here we study the electronic structure and magnetism of the mixed-valent double perovskite YBaCo$_2$O$_6$. We find that YBaCo$_2$O$_6$ is in the formal intermediate-spin (IS) Co$^{3+}$/low-spin (LS) Co$^{4+}$ ground state. The hopping of ${e_g}$ electron from IS-Co$^{3+}$ to LS-Co$^{4+}$ via double exchange gives rise to a ferromagnetic half-metallicity, which well accounts for the recent experiments. The reduction of both magnetization and Curie temperature by oxygen vacancies is discussed, aided with Monte Carlo simulations. We also explore several other possible spin-states and their interesting electronic/magnetic properties. Moreover, we predict that a volume expansion more than 3\%~ would tune YBaCo$_2$O$_6$ into the high-spin (HS) Co$^{3+}$/LS Co$^{4+}$ ferromagnetic state and simultaneously drive a metal-insulator transition. Therefore, spin-states are a useful parameter for tuning the material properties of cobaltates.

\end{abstract}

 \maketitle

\section{Introduction}
Transition metal oxides (TMOs), thanks to the coupling of their lattice, charge, spin, and orbital degrees of freedom, very often display intriguing properties.\cite{tokura2000orbital,dagotto2005complexity,hwang2012emergent,khomskii2014transition,khomskii2020orbital}
Among the various TMOs, perovskite is a very common and important family of materials, possessing abundant functionalities such as ferroelectricity\cite{khomskii2009trend,wang2021chemical}, multiferroicity\cite{wang2015observation}, colossal magnetoresistance\cite{zhang2017giant} and superconductivity.\cite{keimer2015quantum,rubel2014superconducting}
By substituting cations at either $A$ or $B$ site, $AB$O$_3$ perovskite may be ordered into a double perovskite.
Extra complexity including lattice distortion, mixed-charge or mixed-spin state could be introduced by the additional element, giving rise to unique properties such as magnetodielectric effect\cite{yi2015sc2nimno6}, room-temperature magnetoresistance\cite{yamada2019room} and high-temperature magnetic order.\cite{feng2014high}
Compared to $B$-site double perovskite, $A$-site double perovskite $AA'B_2$O$_6$ is much less common.
It is typically found in anion deficient materials, for example, layer-ordered oxygen-deficient perovskite YBaCo$_2$O$_5$\cite{vogt2000low}, which can be oxidized into YBaCo$_2$O$_{5+\delta}$.\cite{akahoshi2001oxygen,khalyavin2007spin}

\begin{figure}[t]
	\centering
	\includegraphics[width=6.5cm]{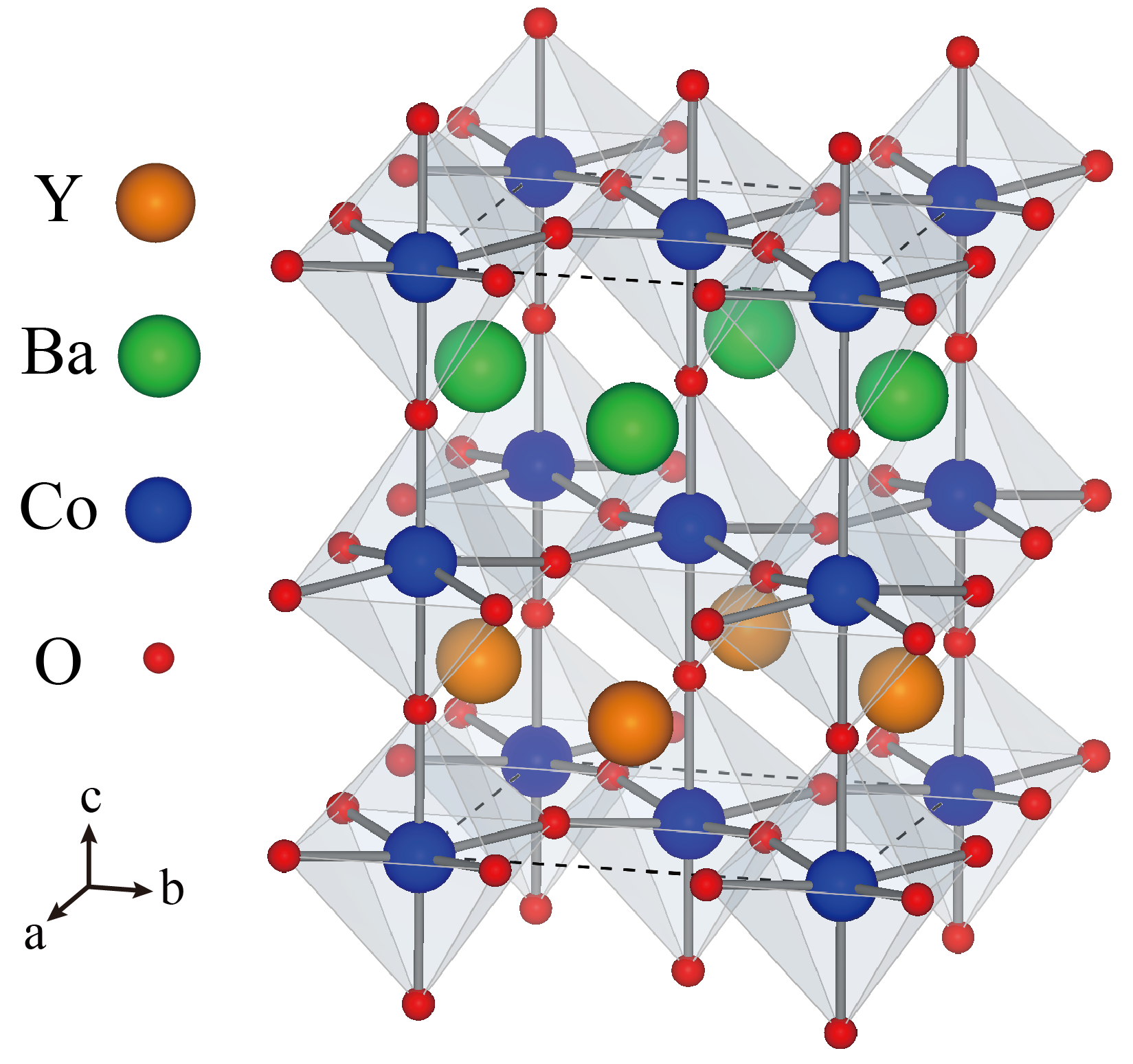}
	\caption{Crystal structure of YBaCo$_2$O$_6$ in a ${\sqrt{2}\times\sqrt{2}\times{1}}$ supercell.}{\label{structure}}
\end{figure}

Arising from a subtle competition between crystal field and Hund exchange, a spin-state issue is intimately related to novel materials behaviors, for instance, magnetic transition and metal-insulator transition.\cite{smith2012evolution,ou2016spin,oka2010pressure}
The importance of spin-states is especially demonstrated in cobaltates since they could be in low-spin (LS), high-spin (HS), or even intermediate-spin (IS) states.
For the prototype material LaCoO$_3$, its thermal spin-state excitations have been extensively studied and ascribed to either an LS-to-HS transition,\cite{haverkort2006spin} LS-to-IS transition,\cite{korotin1996intermediate} or LS-to-(LS+HS)-to-IS transitions\cite{kyomen2005thermodynamical} The complex spin-state transition is still of current interest.\cite{yokoyama2018tensile, tomiyasu2017coulomb} In addition, all different spin-states were actively discussed for $R$BaCo$_2$O$_{5+x}$ with the CoO$_5$ pyramids and CoO$_6$ octahedra.\cite{hu2004different,hu2012spin} Spin-states are also of extensive interest in double perovskites due to their possibly coexistent mixed-charge states.\cite{fan2016unusual,chen2014complete}

Very recently, the $A$-site double perovskite YBaCo$_2$O$_6$ has been successfully synthesized, which contains mixed-valent Co$^{3+}$/Co$^{4+}$.\cite{katayama2018ferromagnetism,goto2018unusual}
It has a unique $A$-site layer-ordered structure with YO-CoO$_2$-BaO-CoO$_2$ stacking sequence along the $c$ axis (see Fig. \ref{structure}).
Two experiments independently obtain YBaCo$_2$O$_6$ by fully oxidizing YBaCo$_2$O$_{5+\delta}$ with different methods, both reporting a ferromagnetic (FM) metallicity with Curie temperature of $T_{\rm{C}}$ = 130 K and magnetization of 1.5 $\mu_{\rm{B}}$/f.u. \cite{katayama2018ferromagnetism} (140 K and 0.8 $\mu_{\rm{B}}$/Co~\cite{goto2018unusual}, i.e., 1.6 $\mu_{\rm{B}}$/f.u.).
However, the spin-state of YBaCo$_2$O$_6$ is still a matter of controversy, and thus the origin of the FM metallicity is not fully understood.
Goto $et$ $al.$ suggested the IS-Co$^{3+}$/IS-Co$^{4+}$ state\cite{goto2018unusual}, while Chikamatsu $et$ $al.$ proposed the HS-Co$^{3+}$/IS-Co$^{4+}$ state.\cite{chikamatsu2021investigation}
Obviously, these controversial spin-states call for an in-depth understanding, and thus we are motivated to study this mixed-valent cobaltate using the spin-state pictures and first-principles calculations.

In this paper, we propose a different spin-state model for YBaCo$_2$O$_6$, based on a detailed investigation of the rich spin-states and the electronic/magnetic properties.
We find that the formal Co$^{3+}$ is in the IS state with $S$ = 1 (3$d^6$, $t_{2g}^5$$e_g^1$) and Co$^{4+}$ in the LS state with $S$ = 1/2 (3$d^5$, $t_{2g}^5$), see Fig. \ref{spin_states}(a). This spin-state yields the FM half-metallic ground state due to the double exchange mechanism, and it provides a good explanation for the experimental FM metallicity. The reduction of both magnetization and Curie temperature by oxygen vacancies is also discussed, aided with Monte Carlo simulations. Moreover, supported by our spin-state analyses and confirmed by first-principles calculations, we predict that with a volume expansion more than 3\%, YBaCo$_2$O$_6$ would undergo a spin-state transition into the HS Co$^{3+}$/LS Co$^{4+}$ state and simultaneously has a metal-insulator transition. Therefore, the spin-states are a useful tuning parameter which is worthy of more attention.

\section{Computational details}
{\label{Computational details}}
First-principles calculations are performed using full-potential augmented plane wave plus local orbital method coded in the WIEN2k package\cite{blaha2020wien2k}.
A ${\sqrt{2}\times\sqrt{2}\times{1}}$ supercell is set for YBaCo$_2$O$_6$, by considering the possible charge order of Co$^{3+}$ and Co$^{4+}$ and different magnetic structures.
Structural optimization is carried out, starting from the experimental lattice parameters $a$ = 3.854 \AA, $b$ = 3.844 \AA, and $c$ = 7.534 \AA\cite{goto2018unusual}.
Atomic relaxation is done using the local-spin-density approximation (LSDA) till all the calculated atomic forces each become smaller than 0.025 eV/\AA.
The muffin-tin sphere radii are chosen to be 2.5, 2.5, 2.0 and 1.2 Bohr for Y, Ba, Co, and O ions, respectively.
The plane-wave cut-off energy for interstitial wave functions is set at 12 Ry, and a mesh of ${5\times5\times5}$ k-points are sampled for integration over the Brillouin zone.
To describe the correlation effect of Co 3$d$ electrons, the LSDA plus Hubbard $U$ (LSDA+$U$) method is employed \cite{Anisimov1991U}, with a common value of $U$ = 6.0 eV and Hund exchange $J_{\rm{H}}$ = 1.0 eV. We also test the $U$ value in a reasonable range of 4-6 eV and $J_{\rm{H}}$ = 0.8 eV, and find that our conclusions remain unchanged.

\section{Results and discussion}
\begin{figure}[]
	\centering
	\includegraphics[width=8.6cm]{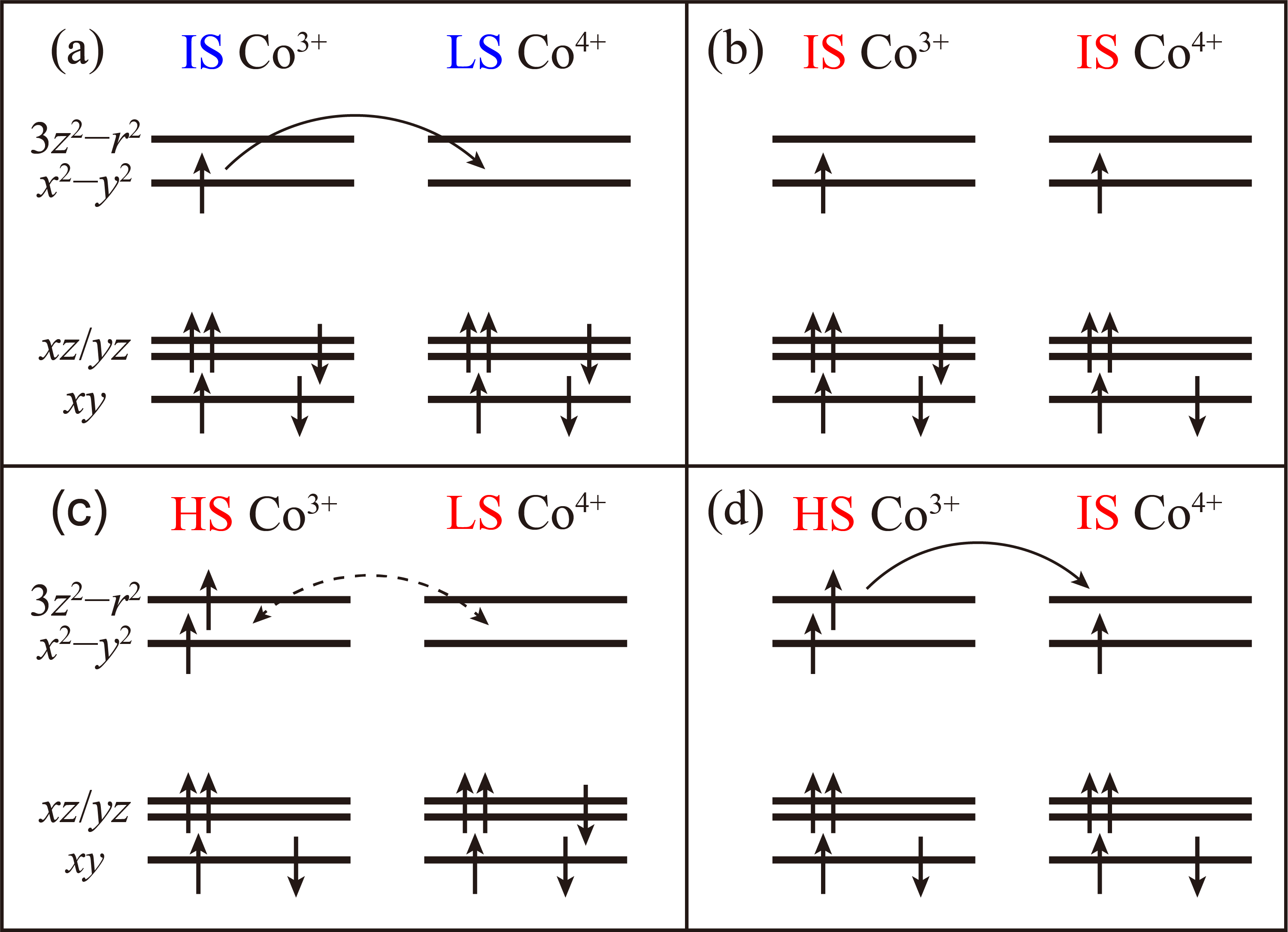}
	\caption{Schematic 3$d$ level diagrams of (a) IS-Co$^{3+}$/LS-Co$^{4+}$, (b) IS-Co$^{3+}$/IS-Co$^{4+}$, (c) HS-Co$^{3+}$/LS-Co$^{4+}$, and (d) HS-Co$^{3+}$/IS-Co$^{4+}$. O 2$p$ orbitals are omitted for brevity.}{\label{spin_states}}
\end{figure}

First we analyze the possible spin-state models for YBaCo$_2$O$_6$ using the spin-state pictures.
Since Co$^{3+}$ has a larger ionic size (and larger Co-O bondlength) than Co$^{4+}$, it ought to have a smaller crystal field splitting and then possibly higher (at least no lower) spin-state, considering a competition between Hund exchange and the crystal field.
LS-Co$^{3+}$/LS-Co$^{4+}$ state has a formal total spin $S$ = 1/2, and the corresponding total moment of 1 $\mu_{\rm{B}}$/f.u. is already much smaller than the experimental one of 1.6 $\mu_{\rm{B}}$/f.u.~\cite{goto2018unusual}. Moreover, this low-spin state together with the small hopping of only the $t_{2g}$ electrons is hard to explain the experimental FM metallicity and therefore is of less concern, which is verified by the following total energy calculations. HS-Co$^{3+}$/HS-Co$^{4+}$ state is also precluded, as it needs a largest lattice volume with a too much cost of elastic energy. In the sense, we will focus on four possible spin-state models for YBaCo$_2$O$_6$ (see Fig. \ref{spin_states}): (a) IS-Co$^{3+}$/LS-Co$^{4+}$, (b) IS-Co$^{3+}$/IS-Co$^{4+}$, (c) HS-Co$^{3+}$/LS-Co$^{4+}$, and (d) HS-Co$^{3+}$/IS-Co$^{4+}$.

A structural optimization is performed for YBaCo$_2$O$_6$ by varying crystal volume from --15\% to +8\% relative to the experimental volume, with a constant lattice parameter ratio. The theoretical equilibrium state shows a volume reduction by 8\%, and thus the lattice parameter is reduced by 2.7\%, which is in line with the common LSDA underestimation of lattice parameters.
Owing to the smaller ionic radius of Y$^{3+}$ (0.90 \AA) than Ba$^{2+}$ (1.35 \AA)\cite{shannon1976revised}, the planar O$^{2-}$ shift to the YO layer by 0.22 \AA.
The asymmetrically distorted CoO$_6$ octahedra have the out-of-plane (in-plane) Co-O bondlengths of 1.83 \AA~ (1.88 \& 1.89 \AA), then the Co $3d$ $x^2-y^2$ level is lower than $3z^2-r^2$, and $xy$ lower than $xz$/$yz$ (see Fig. \ref{spin_states}).

\begin{figure}[]
	\centering
	\includegraphics[width=7.6cm]{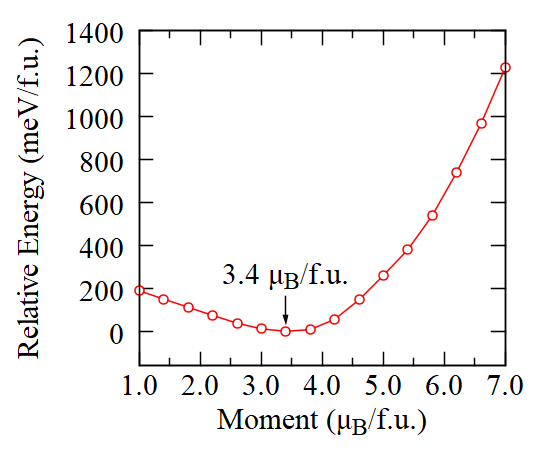}
	\caption{Relative total energies of YBaCo$_2$O$_6$ in fixed-spin-moment calculations by LSDA.}{\label{FSM}}
\end{figure}

To study the electronic structure of YBaCo$_2$O$_6$, we first carry out LSDA calculations using the optimized structure.
The obtained FM metallic solution has a total spin moment of 3.40 $\mu_{\rm{B}}$ per formula unit (f.u.), residing mainly on the two Co ions (1.40$\mu_{\rm{B}}$ each).
This result implies a mixed IS-Co$^{3+}$ ($S$ = 1)/LS-Co$^{4+}$ ($S$ = 1/2) state, which has a total spin $S$ = 3/2, see Fig. \ref{spin_states}(a). To confirm the ground state magnetic moment, we perform fixed-spin-moment (FSM) LSDA calculations, assuming the total spin moments in the range of 1.0-7.0 $\mu_{\rm{B}}$/f.u., see Fig. \ref{FSM}. The obtained results show that the present FM solution with the total moment of 3.4 $\mu_{\rm{B}}$/f.u. is indeed the ground state, and that the LS-Co$^{3+}$/LS-Co$^{4+}$ state with the total moment of 1 $\mu_{\rm{B}}$/f.u. is less stable by about 200 meV/f.u. All other spin-states (e.g., having 5 or 7 $\mu_{\rm{B}}$/f.u., see Figs.\ref{spin_states} and \ref{FSM}) are even much more unstable, let alone the HS-Co$^{3+}$/HS-Co$^{4+}$ state with 9 $\mu_{\rm B}$/f.u. Thus we could infer from these results that YBaCo$_2$O$_6$ is most likely in the IS-Co$^{3+}$/LS-Co$^{4+}$ ground state.

Considering the correlation effect of Co $3d$ electrons, we carry out LSDA+$U$ calculations in which those different possible spin-states of YBaCo$_2$O$_6$ (see Fig. \ref{spin_states}) are initialized by the corresponding occupation number matrix and then fully relaxed in the loops of electronic steps. As seen in Table \ref{table_LSDA+U}, those states are stabilized as they are, except for the IS-Co$^{3+}$/IS-Co$^{4+}$ state which is unstable and converges into the HS-Co$^{3+}$/LS-Co$^{4+}$ state.
These results show that the IS-Co$^{3+}$/LS-Co$^{4+}$ state has the lowest total energy and carries the total moment of 3 ${\rm \mu_B/f.u.}$ as expected for its total spin $S$ = 3/2. In comparison, the HS-Co$^{3+}$/LS-Co$^{4+}$ and HS-Co$^{3+}$/IS-Co$^{4+}$ states have a higher total energy by 114 and 596 meV/f.u., respectively.
These LSDA+$U$ calculations, in line with the above LSDA and FSM calculations, lead to the conclusion that YBaCo$_2$O$_6$ is indeed in the IS-Co$^{3+}$/LS-Co$^{4+}$ ground state.

The Hubbard $U$ = 6 eV used above could be an upper limit for consideration of the Co $3d$ electron correlation. We also test the $U$ value in the reasonable range of 4-6 eV for the LSDA+$U$ calculations, see Table S1 in the Supplemental Information (SI). When $U$ is reduced via 5 eV to 4 eV, the IS-Co$^{3+}$/LS-Co$^{4+}$ ground state remains robust, and all other spin-states become increasingly higher in total energy or even too high to be stable in the LSDA+$U$ calculations (or to say, having a too much higher total energy captured only by LSDA+$U$ FSM calculations). All these LSDA+$U$ ($U$ = 4-6 eV) and LSDA calculations consistently show that YBaCo$_2$O$_6$ is in the IS-Co$^{3+}$/LS-Co$^{4+}$ ground state. Moreover, the increasing stability of the IS-Co$^{3+}$/LS-Co$^{4+}$ ground state against other spin-states with the decreasing $U$ value (from LSDA+$U$ with $U$ = 6 eV to plain LSDA) suggests that the enhanced electron itineracy gains more kinetic energy to stabilize the robust IS-Co$^{3+}$/LS-Co$^{4+}$ ground state, which then produces a FM half-metallic solution as seen below.

Unlike the direct Coulomb repulsion $U$, the Hund exchange $J_{\rm H}$ is actually the difference of the energies of electrons with different spins or orbitals on a same atomic shell, and therefore, $J_{\rm H}$ is almost not screened and not modified when going from an atom to a solid. It is almost a constant for a given element and is typically 0.8-1.0 eV for a $3d$ transition metal\cite{khomskii2014transition}. We therefore have used $J_{\rm H}$ = 1.0 eV for Co $3d$ electrons, and our test calculations using $J_{\rm H}$ = 0.8 eV (see Table S2 in SI) do not affect our conclusions that YBaCo$_2$O$_6$ is in the IS-LS ground state and undergoes a spin-state transition into the HS-LS state within 4\% volume expansion (see more below). We also check the spin-orbit coupling (SOC) effect and find that the SOC does not change the relative stability of the different spin-states but gives some insignificant changes of the magnetic moments (see Table S3 in SI).

\begin{table}[t] 
	\caption{Relative total energies ${\Delta \emph E}$ (meV/f.u.), total and local spin moments (${\rm \mu_B}$) for different spin-states of YBaCo$_2$O$_6$ calculated by LSDA+$U$.}{\label{table_LSDA+U}}
		\begin{tabular*}{0.48\textwidth}{@{\extracolsep{\fill}}lcccc}
			\hline
			spin-states&${\Delta \emph E}$&total&Co$^{3+}$&Co$^{4+}$\\ \hline
			IS-Co$^{3+}$/LS-Co$^{4+}$ & 0 & 3.00 & 1.69 & 1.69 \\ \hline
			IS-Co$^{3+}$/IS-Co$^{4+}$ & \multicolumn{4}{l}{$\rightarrow$ HS-Co$^{3+}$/LS-Co$^{4+}$} \\ \hline
			HS-Co$^{3+}$/LS-Co$^{4+}$ & 114 & 5.00 & 2.80 & 1.70 \\ \hline
			HS-Co$^{3+}$/IS-Co$^{4+}$ & 596 & 5.94 & 2.58 & 2.58 \\ \hline
		\end{tabular*}
\end{table}

\begin{figure}[t]
	\includegraphics[width=8.6cm]{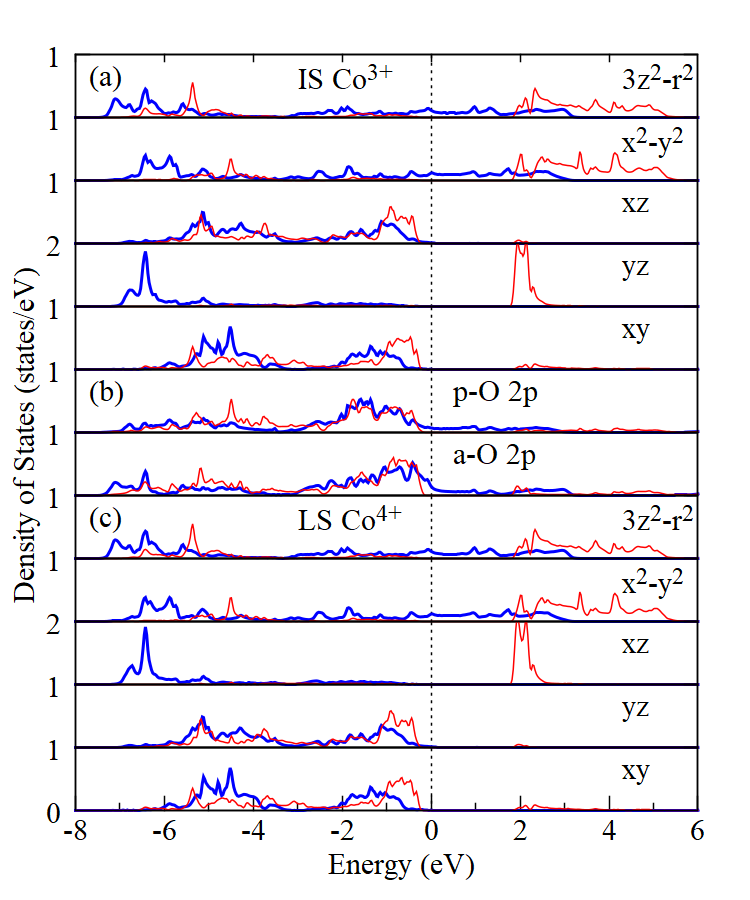}
	\caption{Orbitally resolved DOS of YBaCo$_2$O$_6$ in the formal IS-Co$^{3+}$/LS-Co$^{4+}$ ground state calculated by LSDA+$U$. (a) IS-Co$^{3+}$ 3$d$, (b) planar and apical O 2$p$, and (c) LS-Co$^{4+}$ 3$d$. The blue (red) lines stand for the up (down) spin. The Fermi level is set at zero energy. It is a FM half-metal.}{\label{DOS_IS_LS}}
\end{figure}

To better understand the electronic structure of YBaCo$_2$O$_6$, we plot in Fig. \ref{DOS_IS_LS} the orbitally resolved density of states (DOS) for the IS-Co$^{3+}$/LS-Co$^{4+}$ ground state. The $e_g$ orbitals ($x^2-y^2$ and $3z^2-r^2$) strongly couple to the O $2p$ orbitals and thus form broad $pd\sigma$ bands, and the up-spin channel crosses Fermi level. The $e_g$ bandwidth is much larger than the $e_g$ crystal field splitting associated with the asymmetric distortion of the CoO$_6$ octahedra, and therefore the up-spin $e_g$ orbitals are partially occupied. Moreover, among the $t_{2g}$ electronic levels, the $xy$ is lowest and doubly occupied. The higher $xz$ and $yz$ levels are partially occupied and could form the $yz$/$xz$ orbital order. This ground state solution starts from the formal IS-Co$^{3+}$/LS-Co$^{4+}$ state, and owing to the large $e_g$ hopping (see Figs.\ref{spin_states}(a) and \ref{DOS_IS_LS}), it eventually becomes a mixed and almost homogeneous metallic state (except for a possible $yz$/$xz$ orbital order).
As the up-spin $e_g$ bands cross the Fermi level and the down-spin channel has an insulating gap of 2 eV, YBaCo$_2$O$_6$ displays a clear half-metallicity.

The double exchange mechanism associated with the $e_g$ electron hopping via the intermediate O $2p$ orbitals would favor a FM coupling in the formal IS-Co$^{3+}$/LS-Co$^{4+}$ ground state. To verify this, we carry out LSDA+$U$ calculations for different antiferromagnetic (AFM) states. We find that the A-AFM, C-AFM and G-AFM states lie higher in total energy than the FM ground state by 119, 289 and 328 meV/f.u., respectively, showing both the in-plane and out-of-plane FM couplings. Owing to this FM double exchange, the formal IS-Co$^{3+}$/LS-Co$^{4+}$ ground state becomes an almost homogeneous half-metallic solution, with the Co ions having the same spin moment of 1.69 $\mu_B$. Moreover, the strong $pd\sigma$ hybridization between the Co $3d$ and O $2p$ orbitals, and particularly, the partially occupied up-spin $e_g$ bands give rise to a negative spin polarization of the O ions: a spin moment of --0.07 ${\rm \mu_B}$ (--0.06 ${\rm \mu_B}$) for the planar (apical) O. All these results show that YBaCo$_2$O$_6$ is a FM half-metal due to the double exchange in the formal IS-Co$^{3+}$/LS-Co$^{4+}$ ground state, which well explains the experimental FM metallic behavior~\cite{katayama2018ferromagnetism,goto2018unusual}.

Synthesized by oxidizing YBaCo$_2$O$_{5+x}$\cite{katayama2018ferromagnetism,goto2018unusual,chikamatsu2021investigation}, perfect YBaCo$_2$O$_6$ would have an ideal double perovskite structure, and it has the unusual high valent Co$^{4+}$ ions. Such a high valent oxide often has oxygen vacancies. We now consider the oxygen vacancies and assume YBaCo$_2$O$_{5.9}$, in which 20\% of the otherwise homogeneous Co$^{3.5+}$ ions transform into Co$^{3+}$ ions. Those introduced Co$^{3+}$ ions are in the robust HS state due to the much reduced crystal field in the CoO$_5$ pyramids instead of the original CoO$_6$ octahedra\cite{hu2004different,hu2012spin}. Randomly distributing over the otherwise homogeneous Co$^{3.5+}$ lattice, those HS Co$^{3+}$ ions are AF coupled to their neighboring Co ions due to the oxygen vacancies and the lattice distortion \cite{hu2012spin}. The original FM coupling in the Co$^{3.5+}$ lattice should also be disturbed and reduced by the lattice distortion.

\begin{figure}[t]
	\includegraphics[width=8.6cm]{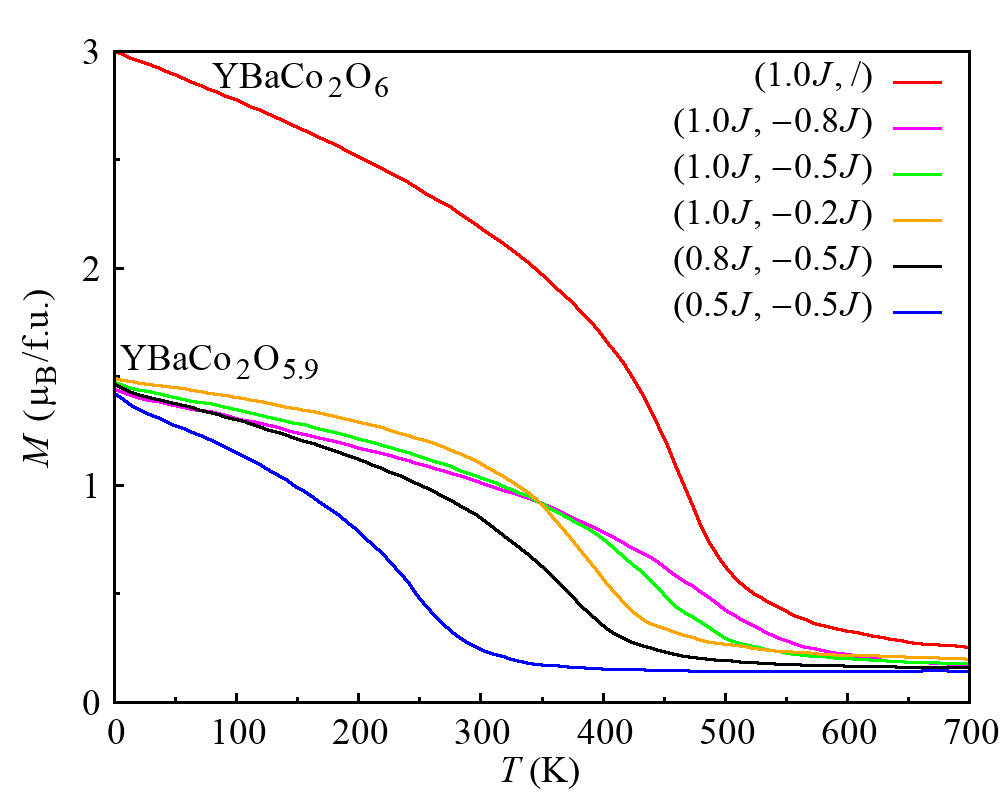}
	\caption{Monte Carlo simulations of the magnetization for YBaCo$_2$O$_6$ and YBaCo$_2$O$_{5.9}$. The nearest neighboring FM coupling $J$ is used for the homogeneous FM metallic ground state of YBaCo$_2$O$_6$ with the average Co$^{3.5+}$ $S$ = 3/4 lattice. Oxygen vacancies in YBaCo$_2$O$_{5.9}$ would yield 20\% Co$^{3+}$ ions which are in the HS state due to the strongly reduced crystal field in the CoO$_5$ pyramids (instead of the CoO$_6$ octahedra). The reduced FM coupling in the host matrix due to the oxygen vacancies and the lattice distortion, and the AF coupling between those HS Co$^{3+}$ ions and their neighbors, as labelled as (0.5$J$, --0.5$J$) for example, both yield a much reduced magnetization and Curie temperature.}{\label{MC}}
\end{figure}

To study those effects of oxygen vacancies on the magnetization and Curie temperature, we carry out Monte Carlo simulations (MCS). We have used a three-dimensional $8\times8\times8$ spin lattice, assuming a Heisenberg spin Hamiltonian with the nearest-neighboring magnetic interactions. Each spin is isotropic and able to rotate arbitrarily in three-dimensional space, and the noncollinearity between the spins has been taken into account. The Metropolis method\cite{metropolis1949monte} is used in the simulations, and the total magnetization is calculated after the system reaches equilibrium at a given temperature. As the formal IS-Co$^{3+}$/LS-Co$^{4+}$ ground state of YBaCo$_2$O$_6$ is an almost homogeneous FM metallic solution, we treat all the Co ions as Co$^{3.5+}$ with the averaged spin = 3/4. Then the FM exchange parameter $J$ = --48.6 meV is estimated from the FM stability over the G-AF state by 328 meV/f.u..

Using the averaged spin = 3/4 and the FM exchange parameter $J$ = --48.6 meV, our MCS show that perfect YBaCo$_2$O$_6$ would have the saturated moment of 3 $\mu_{\rm B}$/f.u. and $T_{\rm{C}}\approx$ 450 K, see Fig. \ref{MC}. When we consider the oxygen vacancies in YBaCo$_2$O$_{5.9}$, we use different pairs of parameters counting the reduced FM coupling in the Co$^{3.5+}$ host lattice and the AF coupling of the introduced HS Co$^{3+}$ ions with their neighbors. We find that all pairs of parameters give almost the same magnetic moment of about 1.5 $\mu_{\rm B}$/f.u., which is in good agreement with the experimental one of 1.5-1.6 $\mu_{\rm B}$/f.u.\cite{katayama2018ferromagnetism,goto2018unusual}. Moreover, the parameters (0.5$J$, --0.5$J$) give $T_{\rm{C}}\approx$ 240 K. All the MCS results show that both the weakened FM coupling in the host and the introduced AF coupling around the oxygen vacancies strongly reduce the $T_{\rm{C}}$ and make it approach the experimental $T_{\rm{C}}$ = 130-140 K\cite{katayama2018ferromagnetism,goto2018unusual}.

\begin{figure}[t]
	\includegraphics[width=8.6cm]{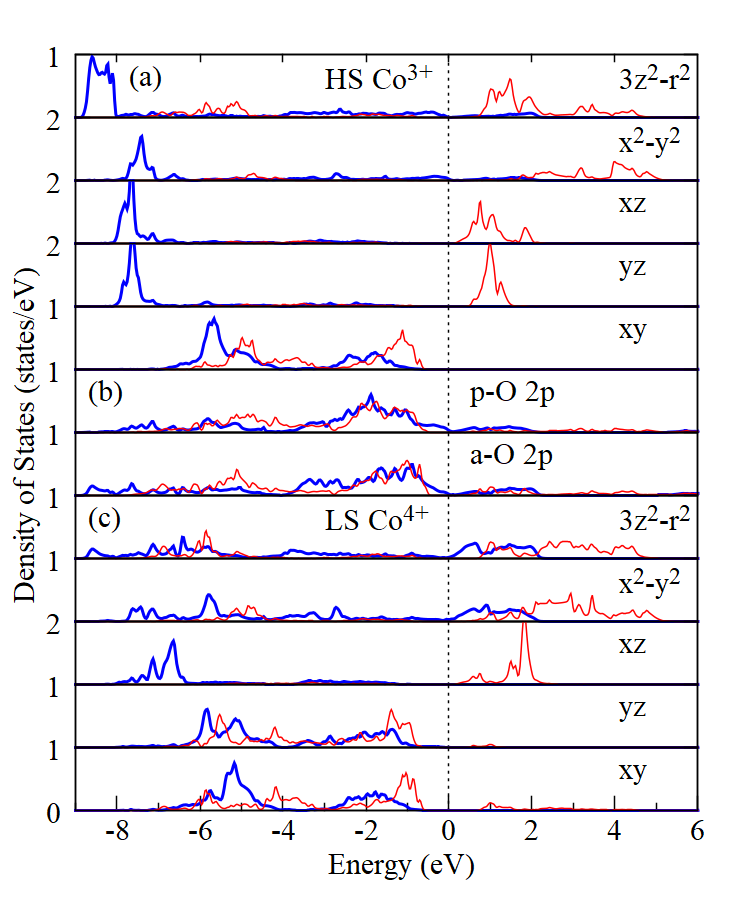}
	\caption{Orbitally resolved DOS of YBaCo$_2$O$_6$ in the HS-Co$^{3+}$/LS-Co$^{4+}$ state. It is a FM insulator.}{\label{DOS_HS_LS}}
\end{figure}

To explore the rich spin-states and the diverse electronic/magnetic properties, we now study the metastable spin-states of YBaCo$_2$O$_6$. We show in Fig. \ref{DOS_HS_LS} the DOS results of the HS-Co$^{3+}$/LS-Co$^{4+}$ state.
Five $3d$ electrons occupy all the up-spin channel of HS-Co$^{3+}$, while the rest one electron resides in the down-spin $xy$ orbital.
For the formal LS-Co$^{4+}$, a clear $t_{2g}^5$ configuration with one hole on $xz$ orbital can be seen.
The total magnetic moment of 5.00 ${\rm \mu_B/f.u.}$ (see Table \ref{table_LSDA+U}) also confirms the HS-Co$^{3+}$ $S$ = 2 and LS-Co$^{4+}$ $S$ = 1/2 states as modeled in Fig. \ref{spin_states}(c).
A finite occupation on the otherwise empty $e_g$ shell of LS-Co$^{4+}$ is mainly due to the Co-O $pd\sigma$ covalence.
And it is the $pd\sigma$ covalence that raises the local spin moment of the LS Co$^{4+}$ to 1.70 ${\rm \mu_B}$ but reduces the HS Co$^{3+}$ moment to 2.80 ${\rm \mu_B}$. One might assume an $e_g$ electron hopping from the HS Co$^{3+}$ to the LS Co$^{4+}$, see Fig. \ref{spin_states}(c). However, such a hopping would change the HS Co$^{3+}$ into IS Co$^{4+}$, and change the LS-Co$^{4+}$ into the IS-Co$^{3+}$, i.e., from the initial spin-state as modeled in Fig. \ref{spin_states}(c) to the final state in Fig. \ref{spin_states}(b). Such a charge fluctuation from the initial $d^6d^5$ to final $d^5d^6$ seems not to change the whole electronic state, but it changes the spin-states a lot. The corresponding Hund exchange energy (by counting each pair of spin-parallel electrons) changes from 10$J_{\rm{H}}$ + 4$J_{\rm{H}}$ to 7$J_{\rm{H}}$ + 6$J_{\rm{H}}$, i.e., there would be an energy cost of one $J_{\rm{H}}$ about 1 eV associated with this charge fluctuation. This large energy cost would make the so-called final IS-Co$^{3+}$/IS-Co$^{4+}$ state unstable, which indeed converges to the metastable initial HS-Co$^{3+}$/LS-Co$^{4+}$ state in our LSDA+$U$ calculations (see Table I). Therefore, this spin blockade mechanism\cite{chang2009spin} makes the $e_g$ electron hopping virtual and the HS-Co$^{3+}$/LS-Co$^{4+}$ state is insulating as seen in Fig. \ref{DOS_HS_LS}.

However, this virtual $e_g$ electron hopping (through O 2$p$) would yield a FM superexchange, even in this insulating solution, to maximize the Hund exchange in the virtually excited intermediate state, see Fig. \ref{spin_states}(c). Our LSDA+$U$ calculations show that for this insulating HS-Co$^{3+}$/LS-Co$^{4+}$ solution, the FM state is more stable than the G-AFM state by 308 meV/f.u. This result is well comparable with the above one of 328 meV/f.u. for the FM stability of the half-metallic IS-Co$^{3+}$/LS-Co$^{4+}$ ground state against the G-AFM state. Note that normally a localized FM superexchange is weaker than an itinerant FM exchange, but that the HS-Co$^{3+}$/LS-Co$^{4+}$ state has a larger total spin ($S$ = 5/2) than the IS-Co$^{3+}$/LS-Co$^{4+}$ ground state with a total spin of 3/2. These results show that the HS-Co$^{3+}$/LS-Co$^{4+}$ state would be an interesting FM insulator with a pretty strong FM superexchange, although it is not the ground state for YBaCo$_2$O$_6$ and is a contrast with the experimental metallic behavior\cite{katayama2018ferromagnetism,goto2018unusual}.

\begin{figure}[t]
	\includegraphics[width=8.6cm]{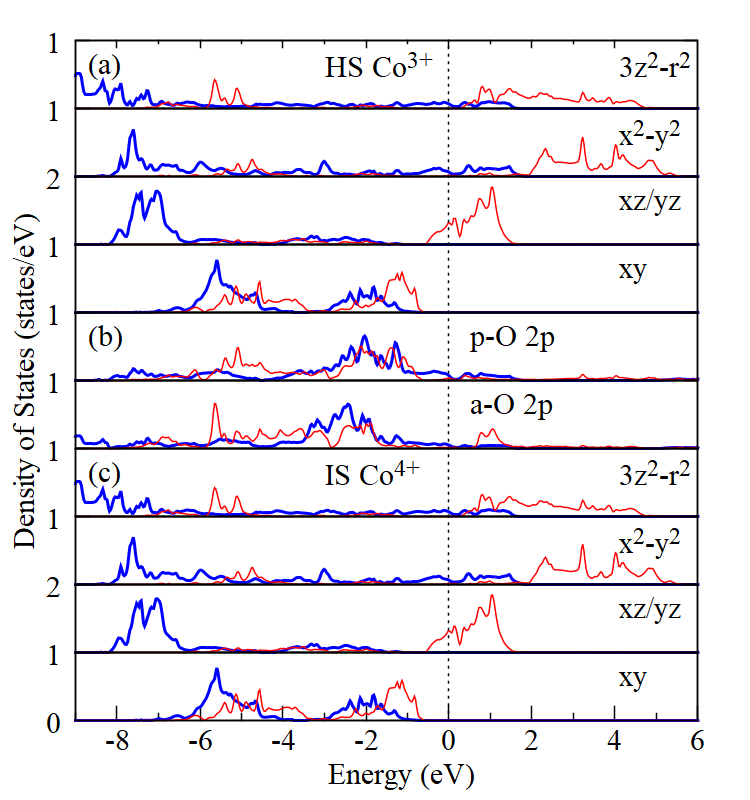}
	\caption{Orbitally resolved DOS of YBaCo$_2$O$_6$ in the HS-Co$^{3+}$/IS-Co$^{4+}$ state. It is a FM metal.}{\label{DOS_HS_IS}}
\end{figure}

As for another metastable HS-Co$^{3+}$/IS-Co$^{4+}$ state [Fig. \ref{spin_states}(d)], we plot its DOS results in Fig. \ref{DOS_HS_IS}.
A double exchange occurs in the $e_g$ shell which shows similarity to the IS-Co$^{3+}$/LS-Co$^{4+}$ state but has more electrons to participate.
Hence, the up-spin broad $e_g$ bands cross the Fermi level, giving a homogeneous metallic solution with the same local moment of 2.58 ${\rm \mu_B}$ for both the formal Co$^{3+}$ and Co$^{4+}$ (see Table \ref{table_LSDA+U}).
Owing to the strong real hopping between the Co ions, associated with the double exchange, and their large magnetic moments, the HS-Co$^{3+}$/IS-Co$^{4+}$ state would have an even stronger FM coupling.
Nevertheless, the HS-Co$^{3+}$/IS-Co$^{4+}$ state may be hard to achieve, because their 3$d$ electrons are more expansive in space than
the IS-Co$^{3+}$/LS-Co$^{4+}$ and the large lattice elastic energy cost gives a much higher total energy (596 meV/f.u.) than the IS-Co$^{3+}$/LS-Co$^{4+}$ ground state. Note that the broad $e_g$ bands raise the Fermi level so that it also strides over the down-spin $xz$/$yz$ bands, making the formal HS-Co$^{3+}$/IS-Co$^{4+}$ state metallic (not half-metallic) and reducing the total spin moment from the expected 7 ${\rm \mu_B}$/f.u. to 5.94 ${\rm \mu_B}$/f.u.. These results show that YBaCo$_2$O$_6$, if stabilized into the formal HS-Co$^{3+}$/IS-Co$^{4+}$ state, would be a strong FM metal (or even possibly a half-metal, see Fig. S2 in SI).

\begin{figure}[]
	\includegraphics[width=8.6cm]{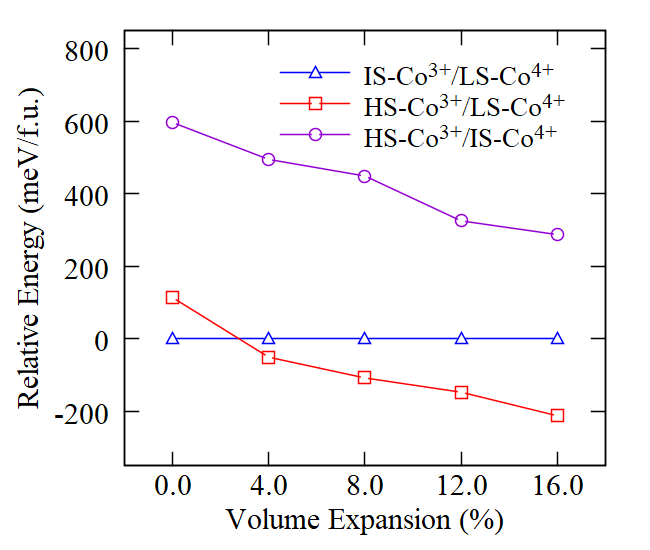}
	\caption{LSDA+$U$ total energies of YBaCo$_2$O$_6$ with a volume expansion in different spin-states. A 3\% expansion would trigger a spin-state transition from IS-Co$^{3+}$/LS-Co$^{4+}$ to HS-Co$^{3+}$/LS-Co$^{4+}$.}{\label{V-Change}}
\end{figure}

Apparently, a cobaltate such as YBaCo$_2$O$_6$ has rich spin-states and diverse electronic/magnetic properties, and therefore, we are motivated to find a way to fine tuning of the spin states for possible functionality/applications. As the spin-states are sensitive to the crystal field, here we simulate the lattice volume expansion to check the switch of the spin-states and possible metal-insulator transition and/or magnetic transition. In practice, a lattice expansion may be induced by cation substitution, heating, and tensile strain, etc. As seen in Fig. \ref{V-Change}, with the increasing lattice volume, the energy difference between the HS-Co$^{3+}$/IS-Co$^{4+}$ and IS-Co$^{3+}$/LS-Co$^{4+}$ states decreases but the former is still much higher than the later in total energy due to the large elastic energy cost. Thus the HS-Co$^{3+}$/IS-Co$^{4+}$ state seems to need an extraordinary large volume to stabilize itself. In strong contrast, when the volume expansion is more than 3\%, the HS-Co$^{3+}$/LS-Co$^{4+}$ state with an average medium ionic size would have lower total energy than the IS-Co$^{3+}$/LS-Co$^{4+}$ state and then turn into the ground state. This HS-Co$^{3+}$/LS-Co$^{4+}$ state has a strong FM coupling but is insulating as discussed above. By a comparison with the FM half-metallic IS-Co$^{3+}$/LS-Co$^{4+}$ state, we predict that a volume expansion of YBaCo$_2$O$_6$ more than 3\% could trigger an interesting spin-state transition from IS-Co$^{3+}$/LS-Co$^{4+}$ to HS-Co$^{3+}$/LS-Co$^{4+}$ and simultaneously a metal-insulator transition, but that surprisingly the FM order persists. This prediction could be worthy of an experimental study.

\section{Conclusions}
In summary, cobaltates have varying and tunable spin-states and diverse properties. Here we study the $A$-site ordered double perovskite YBaCo$_2$O$_6$, using the rich spin-state pictures, magnetic coupling analyses, first-principles calculations, and Monte Carlo simulations. We propose a different spin-state model, i.e., the formal IS-Co$^{3+}$/LS-Co$^{4+}$ state, and demonstrate that it can well explain the experimental FM metallicity. The $e_g$ electron hopping via the double exchange mechanism could even create an ideal FM half-metallicity. A reduction of both the magnetization and Curie temperature due to oxygen vacancies is also discussed. Moreover, we explore several other possible spin-states and their interesting electronic/magnetic properties. In particular, we predict that upon a volume expansion more than 3\%, this cobaltate could undergo a spin-state transition from IS-Co$^{3+}$/LS-Co$^{4+}$ to HS-Co$^{3+}$/LS-Co$^{4+}$, and simultaneously transit from a FM half-metal to a FM insulator. This work highlights the rich spin-states which are a useful parameter for tuning the materials properties of cobaltates.

\section{Acknowledgements}
{\label{Acknowledgements}}
This work was supported by the NSF of China (Grant No.11674064) and by the National Key Research and Development
Program of China (Grant No. 2016YFA0300700).

\bibliographystyle{apsrev4-1}
\bibliography{YBaCo2O6}

\newpage
\clearpage

\onecolumngrid
\begin{appendix}
	\setcounter{figure}{0}
	\setcounter{table}{0}
	\renewcommand{\thefigure}{S\arabic{figure}}
	\renewcommand{\thetable}{S\arabic{table}}
	\renewcommand{\tablename}{Table}
	\renewcommand{\figurename}{Fig.}

\subsection*{Supporting Information for ``Ferromagnetic half-metallicity in YBaCo$_2$O$_6$ and spin-states driven metal-insulator transition''}

 \begin{table*}[h] 
\normalsize
	\caption{Relative total energies ${\Delta \emph E}$ (meV/f.u.) for different spin-states of YBaCo$_2$O$_6$ calculated by LSDA+$U$ with different $U$ values. The total magnetic moment per formula unit (${\rm \mu_B/f.u.}$) is listed in the round brackets. Fixed-spin-moment (FSM) results in LSDA+$U$ are also included.}{\label{table_S1}}
	\begin{tabular}{c@{\hskip3mm}c@{\hskip4mm}c@{\hskip4mm}c@{\hskip4mm}c@{\hskip4mm}}
		\hline\hline
		&IS-Co$^{3+}$/LS-Co$^{4+}$&IS-Co$^{3+}$/IS-Co$^{4+}$&HS-Co$^{3+}$/LS-Co$^{4+}$&HS-Co$^{3+}$/IS-Co$^{4+}$\\ \hline
		$U$ = 6 eV & 0 (3.00 ${\rm \mu_B}$) & $\rightarrow$ HS/LS & 114 (5.00 ${\rm \mu_B}$) & 596 (5.94 ${\rm \mu_B}$) \\
		&  &  & & 770 (FSM, 7.00 ${\rm \mu_B}$) \\ \hline
		$U$ = 5 eV & 0 (3.00 ${\rm \mu_B}$) & $\rightarrow$ HS/LS & 190 (4.91 ${\rm \mu_B}$) & $\rightarrow$ IS/LS \\
		&  &  &  & 970 (7.00 ${\rm \mu_B}$) \\ \hline
		$U$ = 4 eV & 0 (3.00 ${\rm \mu_B}$) & $\rightarrow$ IS/LS & $\rightarrow$ IS/LS & $\rightarrow$ IS/LS \\
		&  &  & 301 (5.00 ${\rm \mu_B}$) & 1138 (7.00 ${\rm \mu_B}$) \\
		\hline\hline
	\end{tabular}
\end{table*}

\begin{table*}[h] 
\normalsize
	\caption{Relative total energies ${\Delta \emph E}$ (meV/f.u.) of the IS-LS and HS-LS states of YBaCo$_2$O$_6$ with different $J_{\rm H}$.}{\label{table_S2}}
	\begin{tabular}{c@{\hskip3mm}c@{\hskip4mm}c@{\hskip4mm}c@{\hskip4mm}}
		\hline\hline
		volume expansion&spin-states& $J_{\rm H}$ = 1 eV& $J_{\rm H}$ = 0.8 eV\\ \hline
		\multirow{2}{*}{+0\%} & IS-Co$^{3+}$/LS-Co$^{4+}$ & 0 & 0 \\
		& HS-Co$^{3+}$/LS-Co$^{4+}$ & 114 & 76 \\ \hline
		\multirow{2}{*}{+4\%} & IS-Co$^{3+}$/LS-Co$^{4+}$ & 47 & 85 \\
		& HS-Co$^{3+}$/LS-Co$^{4+}$ & 0 & 0 \\
		\hline\hline
	\end{tabular}
\end{table*}

\begin{table*}[h] 
\normalsize
	\caption{Relative total energies ${\Delta \emph E}$ (meV/f.u.), local spin and orbital moments (${\rm \mu_B}$/Co) of the IS-LS and HS-LS states of YBaCo$_2$O$_6$ calculated by LSDA+$U$ and LSDA+$U$+SOC.}{\label{table_S3}}
	\begin{tabular}{c@{\hskip3mm}c@{\hskip4mm}c@{\hskip4mm}c@{\hskip1mm}c@{\hskip4mm}c@{\hskip4mm}c@{\hskip4mm}c@{\hskip4mm}}
		\hline\hline
		\multirow{2}{*}{volume expansion}&\multirow{2}{*}{spin-states}& \multicolumn{2}{c}{LSDA+$U$} & & \multicolumn{3}{c}{LSDA+$U$+SOC} \\ \cline{3-4} \cline{6-8}
		& & ${\Delta \emph E}$ & $M_{spin}$ & &${\Delta \emph E}$ & $M_{spin}$ & $M_{orb}$ \\ \hline
		\multirow{2}{*}{+0\%} & IS-Co$^{3+}$/LS-Co$^{4+}$ & 0 & 1.69/1.69 & & 0 & 1.67/1.67 & 0.29/0.30 \\
		& HS-Co$^{3+}$/LS-Co$^{4+}$ & 114 & 2.80/1.70 & & 114 & 2.72/1.67 & 0.16/0.36 \\ \hline
		\multirow{2}{*}{+4\%} & IS-Co$^{3+}$/LS-Co$^{4+}$ & 47 & 1.78/1.78 & & 48 & 1.75/1.75 & 0.33/0.33 \\
		& HS-Co$^{3+}$/LS-Co$^{4+}$ & 0 & 2.90/1.51 & & 0 & 2.86/1.49 & 0.18/0.28 \\
		\hline\hline
	\end{tabular}
\end{table*}

\begin{figure*}[h]
	\includegraphics[width=11cm]{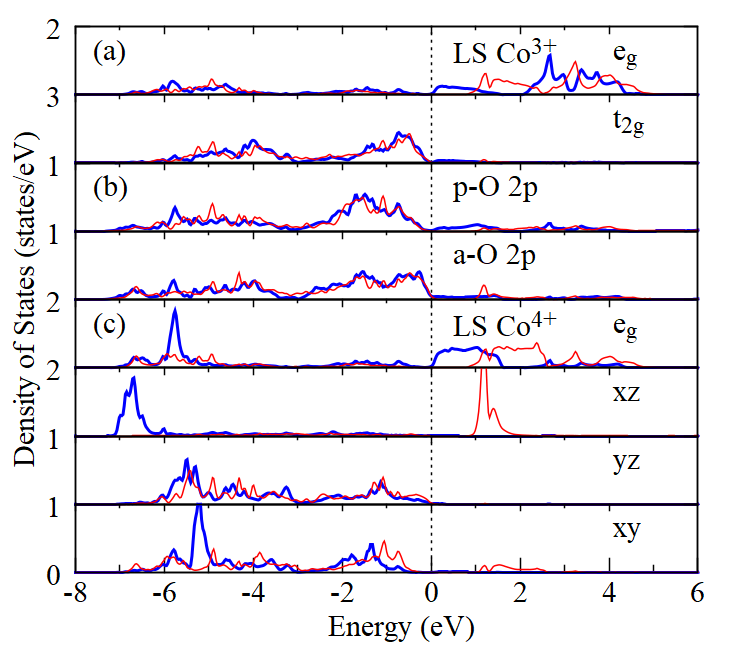}
	\caption{DOS of YBaCo$_2$O$_6$ in the LS-Co$^{3+}$/LS-Co$^{4+}$ state by LSDA+$U$. It is a paramagnetic insulator.}{\label{DOS_LS_LS}}
\end{figure*}

\begin{figure*}[h]
	\includegraphics[width=11cm]{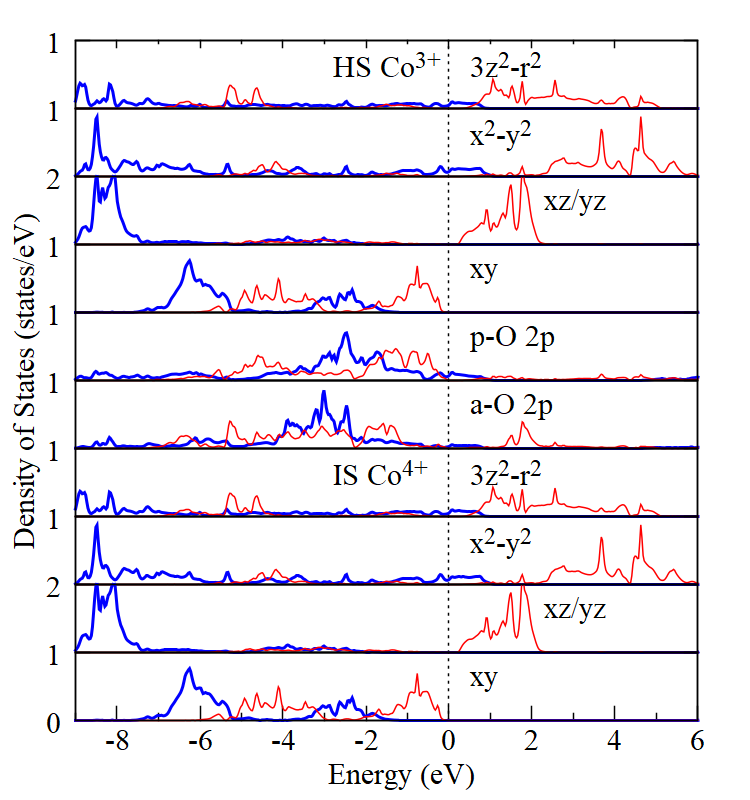}
	\caption{DOS of YBaCo$_2$O$_6$ in the HS-Co$^{3+}$/IS-Co$^{4+}$ state by LSDA+$U$ FSM calculation with $M_{tot}$ = 7.00 ${\rm \mu_B}$/f.u.. It is a FM half-metal.}{\label{DOS_HS_IS_FSM}}
\end{figure*}

\end{appendix}

\end{document}